\documentclass{article}

\usepackage{spconf,amsmath,amsfonts,graphicx}
\usepackage{amssymb,textcomp,mathtools}
\usepackage{bm,upgreek,algorithm,hyperref}
\usepackage{multirow,booktabs,hhline,array}
\usepackage{cite,url,makecell,setspace}
\usepackage{xcolor}
\pdfoutput=1

\def\thline{\noalign{\hrule height 1.0pt}}

\renewcommand{\vec}[1]{\bm{\mathrm{#1}}}

\title{Rethinking the Separation Layers in Speech Separation Networks}
\name{Yi~Luo$^{\dag}$, Zhuo~Chen$^\mathsection$, Cong~Han$^{\dag}$, Chenda~Li$^\ddagger$, Tianyan~Zhou$^\mathsection$, Nima~Mesgarani$^{\dag}$}
\address{
  $^{\dag}$Department of Electrical Engineering, Columbia University \\
  $^\ddagger$Department of Computer Science and Engineering, Shanghai Jiao Tong University\\
  $^\mathsection$Microsoft Corporation
}

\begin{document}
\ninept
\maketitle

\begin{abstract}
Modules in all existing speech separation networks can be categorized into single-input-multi-output (SIMO) modules and single-input-single-output (SISO) modules. SIMO modules generate more outputs than input, and SISO modules keep the numbers of input and output the same. While the majority of separation models only contain SIMO architectures, it has also been shown that certain two-stage separation systems integrated with a post-enhancement SISO module can improve the separation quality. Why performance improvements can be achieved by incorporating the SISO modules? Are SIMO modules always necessary? In this paper, we empirically examine those questions by designing models with varying configurations in the SIMO and SISO modules. We show that comparing with the standard SIMO-only design, a mixed SIMO-SISO design with a same model size is able to improve the separation performance especially under low-overlap conditions. We further validate the necessity of SIMO modules and show that SISO-only models are still able to perform separation without sacrificing the performance. The observations allow us to rethink the model design paradigm and present different views on how the separation is performed.
\end{abstract}

\section{Introduction}
\label{sec:introduction}
Speech separation aims at separating one or all active speakers from a given mixture. Tremendous efforts have been made by the community in exploring better problem formulations, model designs, training objectives, and data configurations, with the advances in deep neural networks \cite{hershey2016deep, yu2017permutation, luo2018tasnet, wang2018alternative, le2019sdr, drude2019sms, wichern2019wham, cosentino2020librimix, chen2020continuous}.  Given the mixture signal is usually considered as a single input, separation models can be broadly categorized into single-input-single-output (\textit{SISO}) systems and single-input-multi-output 
(\textit{SIMO}) systems. As its name suggests, a SISO system usually consists a stack of one to one mapping layers, extracting one speaker from the mixture at each time. SISO networks are typically designed for guided source separation (GSS) or speech enhancement tasks \cite{vincent2014blind, wang2018voicefilter, chen2018multi, kanda2019guided, vzmolikova2019speakerbeam, xu2019time}, where a bias is often needed to distinguish the target speaker. When there are more than one source that need to be estimated, the single output separation needs to be performed multiple times, one for each source. 
In contrast, the SIMO systems are the standard design for blind source separation (BSS) \cite{kolbaek2017multitalker, luo2019conv, stoller2018wave, zeghidour2020wavesplit, nachmani2020voice}. The SIMO system targets at separating $C$ sources simultaneously. To fulfill this task, on top of the one to one mapping layers, there always exists one or more one-to-many mapping layers that convert the single input signal to multiple source output, i.e. one to many mapping. For example, in standard masking-based BSS models, $C$ masks are generally estimated from the last layer in the network. In certain iterative separation methods, two masks are estimated from the last layer in the model representing one target source and the residual signal, respectively \cite{takahashi2019recursive}. In \cite{chang2019mimo}, though named as ``MIMO'' network, the system is essentially a SIMO system where the input feature consist of multi-microphone information.

%  From the differences in the single-output and multi-output systems, we can further split any model into single-input-multi-output \textit{(SIMO)} modules and single-input-single-output \textit{(SISO)} modules. Both modules can contain multiple layers to construct a deeper architecture. A SIMO module performs separation as it generates more outputs than input, and a SISO module can be treated as a complex transformation or a form of enhancement since it assumes a same number of input and output. In other words, the SIMO module defines the \textit{separation layers} in a separation network. Almost all separation systems are SIMO-only, where a single feed-forward layer serving as the output layer of the entire model maps the input into multiple outputs. 
Literatures have also explored the combination of SIMO and SIMO architecture for further performance improvement. A commonly applied integration is to use a SISO network for post-enhancement module on the output of the SIMO separation result \cite{kolbaek2017multitalker, wichern2019wham, delfarah2019deep, yoshioka2019low}, while typically the two modules are not jointly optimized. However, as the combination systems usually contain a significantly larger parameter size which could also results in potential performance improvement for a pure SIMO network, little is known about the roles of SIMO and SISO modules in a BSS separation system. Why performance improvements can be achieved by incorporating the SISO modules? For a given model size, how to properly arrange the sizes of SIMO and SISO modules to achieve a best performance? Are SIMO modules always necessary? 

In this paper, we empirically analyze different model configurations, including the standard SIMO-only model, the mixed SIMO-SISO models, and the SISO-only models, on their effectiveness on the separation performance. The SIMO-SISO models follow the design of a pre-separation and post-enhancement pipeline. Unlike the SISO designs in GSS tasks (e.g. speaker extraction), the SISO-only models here do not use external bias information but perform separation in an iterative way. We explore various hyperparameter configurations all under a same total model size, and observe that the mixed SIMO-SISO design is able to improve the performance especially under low-overlap conditions. Such observation indicates that the SISO modules are particularly beneficial in single-speaker regions. We also validate that SISO-only models are still able to perform separation without sacrificing the performance, challenging the role of SIMO modules and the general problem formulation of speech separation. Our results not only allow us to rethink the model design paradigm, but also present different views on the general framework of BSS.

The rest of the paper is organized as follows. Section~\ref{sec:separate} introduces the SIMO-only, the mixed SIMO-SISO modules, and the SISO-only modules for the BSS task. Section~\ref{sec:config} summarizes the experiment configurations. Section~\ref{sec:result} presents the experiment results and discussions. Section~\ref{sec:conclusion} concludes the paper.

\section{Configurations of SIMO and SISO modules in separation networks}
\label{sec:separate}
\begin{figure*}[!ht]
	\small
	\centering
	\includegraphics[width=2\columnwidth]{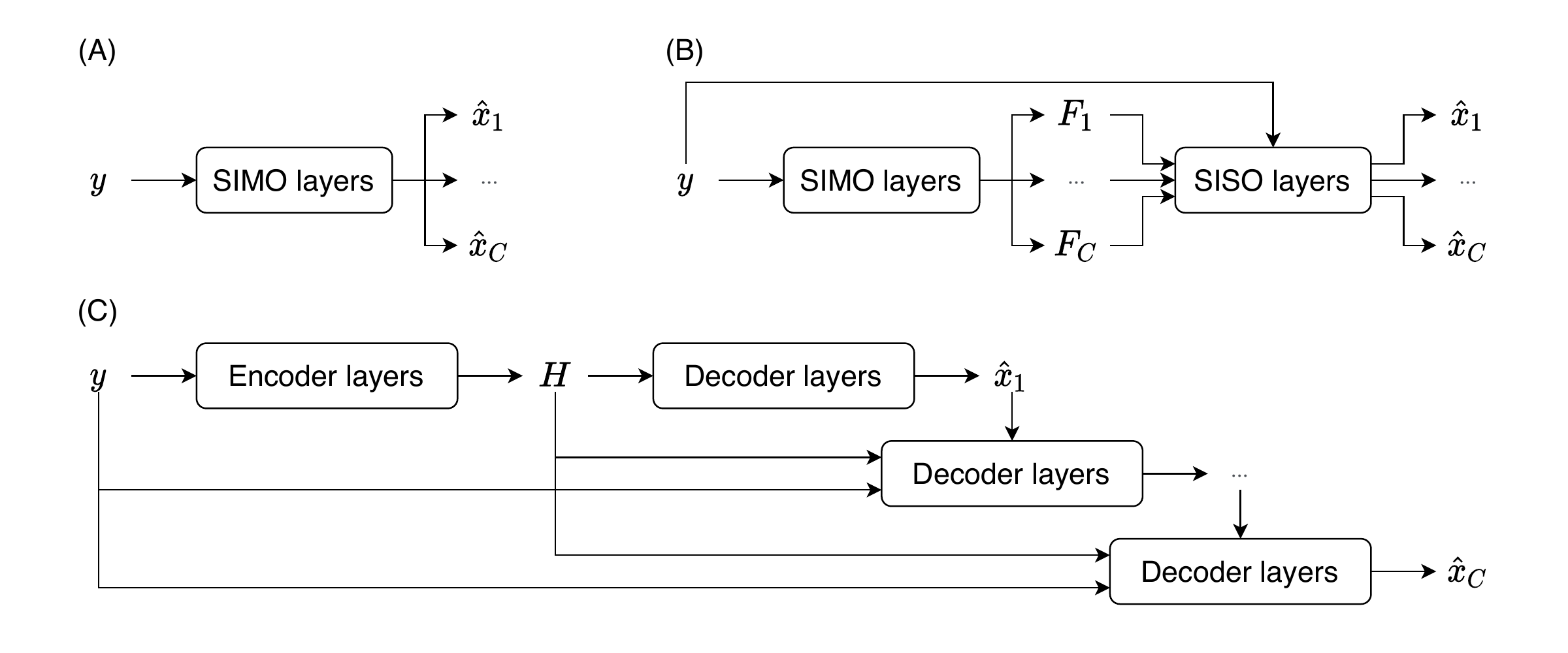}
	\caption{Flowchart of different configurations on the separation models. (A) Standard separation model with a single SIMO module that estimates $C$ target sources. (B) A SIMO module first generates $C$ intermediate features for the $C$ sources, and a SISO module takes each of the feature as input and estimates the target sources. (C) A SISO encoder module first generates one intermediate feature from the input mixture. A SISO decoder module takes the feature as input and estimates the first target source. The input mixture, intermediate feature, and the target source are passed again to the decoder module to estimate the second target source. Such procedure is repeated until all sources are separated.}
	\label{fig:flowchart}
\end{figure*}

\subsection{Problem formulation and backbone architecture}

We start this section with the problem formulation and the backbone architecture used in all configurations. We adopt the single-channel blind separation formulation in this paper, and the multi-channel case can be easily extended.

A mixture signal $\vec{y} \in \mathbb{R}^{1\times t}$ containing $C$ speakers $\{\vec{x}_i\}_{i=1}^C$  and an optional noise $\vec{n}$ is represented as:
\begin{align}
    \vec{y} = \sum_{i=1}^C \vec{x}_i + \vec{n}
\end{align}

The time-domain audio separation network (TasNet) equipped with dual-path RNN (DPRNN) \cite{luo2019dual} is used as the backbone model. We assume that in all configurations the number of DPRNN blocks is fixed to $M$ across all configuration for a fair comparison. $C$ outputs $\{\hat{\vec{x}}_i\}_{i=1}^C$ are estimated from $\vec{y}$ by the DPRNN-TasNet. We denote the linear mapping function defined by the waveform encoder in the DPRNN-TasNet as $\mathcal{E}(\cdot)$, and we omit the waveform encoder and decoder in Figure~\ref{fig:flowchart} for the sake of simplicity.

\subsection{SIMO-only model design}

The SIMO-only model design is the default design of almost all current separation models. Figure~\ref{fig:flowchart} (A) shows the flowchart for the design. The $M$ DPRNN blocks all belong to the SIMO module, and the $C$ targets are estimated from the output layer of the module, which is typically a fully-connected (FC) layer with $C$ output heads. This is also the original design for the DPRNN-TasNet.

\subsection{Mixed SIMO-SISO model design}

Figure~\ref{fig:flowchart} (B) illustrates the flowchart for the mixed SIMO-SISO design. The $M$ DPRNN blocks are split into a SIMO module and a SISO module, where each module contains $K$ and $M-K$ blocks, respectively. Similar to the SIMO-only design, the SIMO module is first applied on the input mixture $\vec{y}$ to create $C$ intermediate features $\{\vec{F}_i\}_{i=1}^C \in \mathbb{R}^{N\times L}$. Each of the intermediate feature $\vec{F}_i$, together with the encoder output of the mixture signal $\mathcal{E}(\vec{y})$, are then passed to the SISO module, which is shared by all SIMO output features, to generate the final estimations $\{\hat{\vec{x}}_i\}_{i=1}^C$. The two modules are jointly optimized and no extra training objective is applied to the intermediate features.

Note that unlike the SIMO-only design where the outputs of the SIMO module are typically $C$ masks applied to the input mixture, here the output layer for the SIMO module can simply be a linear FC layer and the outputs do not need to be applied to the mixture. We also test the setting where the outputs from the SIMO module are indeed the $C$ masks and use the masked mixture encoder output as $\{\vec{F}_i\}_{i=1}^C$, and add $\{\vec{F}_i\}_{i=1}^C$ to the SISO outputs to form the final separated sources. This matches the standard pipeline in pre-separation and post-enhancement models, where the SIMO module servers as the pre-separation module and the SISO module is the post-enhancement module. We empirically find that such setting leads to identical performance as the simpler pipeline in Figure~\ref{fig:flowchart} (B).

\subsection{SISO-only model design}

Figure~\ref{fig:flowchart} (C) presents the flowchart for the SISO-only design. Since no SIMO module is present in the entire model, iterative separation has to be applied in order to separate all $C$ targets. We split the $M$ layers in the SISO module into $K$ \textit{encoder layers} and $M-K$ \textit{decoder layers}, where the encoder layers are applied only once and the decoder layers are applied in every iteration. In other words, the encoder layers map the mixture into a latent representation shared by all iterations, and the decoder layers separate different targets based on the representation.

The mixture $\vec{y}$ is passed to the encoder layers to generate one sequence of intermediate feature $\vec{H} \in \mathbb{R}^{N\times L}$. In the first iteration, the encoder output of the mixture $\mathcal{E}(\vec{y})$, the intermediate feature $\vec{H}$, and an all-zero feature with the same shape as $\mathcal{E}(\vec{y})$ are passed to decoder layers to generate the first output $\hat{\vec{x}}_1$. In the $j$-th iteration where $j>1$, $\mathcal{E}(\vec{y})$, $\vec{H}$ and the encoder output of the residual signal $\mathcal{E}(\vec{y} - \sum_{k=1}^{j-1} \hat{\vec{x}}_k)$ are passed to decoder layers to generate the $j$-th output $\hat{\vec{x}}_j$. Note that here we assume that the number of target sources is known in advance, but the same procedure can also be applied in the task of separating unknown number of speakers.

\subsection{Discussions}

In the mixed SIMO-SISO design, both the intermediate feature and the input mixture are sent to the SISO module. We make this design because empirically we find that it leads to better performance than simply sending the intermediate feature to the SISO module. Similarly, in the SISO-only design we also send all available features to the decoder layers. Why such feature fusion leads to constantly better performance is left for future work, and in this paper we mainly focus on the variations on the model design paradigm.

The iterative SISO-only design can be connected to the GSS framework, where the bias information comes from the residual signal in the previous iteration. The main difference here is that in GSS frameworks the bias information is typically related to the target to be extracted, e.g. speaker-related feature or content-related feature, while in the SISO-only design the bias information is related to all the signals that have not been separated. There could be other configurations of feature fusion, e.g. using all the separated signals instead of the residual signal as the bias, but we leave it for future work to validate.

Also note that in a recent literature, a newly proposed training method, the serialized output training (SOT) \cite{kanda2020serialized}, applies the SISO-only configuration without iterative separation. SOT is designed for multi-talker automatic speech recognition (ASR), and it concatenates all target output sequences in to a single sequence as the training target. Together with an encoder-decoder architecture, the decoder sequentially generates the predicted labels for all speakers. Although SOT can also be extended to the task of speech separation, one main difference between ASR and separation is that the length of the output sequences in separation tasks is always the same as the input, while the length of the output sequences in ASR tasks can vary for different speakers. Such generative decoding mechanism might have trouble in the separation outputs as the total length of the output sequence can be significantly longer than that in ASR tasks.
% In this paper we only consider the iterative separation mechanism in the SISO-only design, and leave other possible mechanisms for future work.

\section{Experiment configurations}
\label{sec:config}
\subsection{Dataset}

We evaluate the different model configurations on a simulated noisy reverberant two-speaker dataset \cite{luo2020end}. 20000, 5000 and 3000 4-second long utterances are simulated for training, validation and test sets, respectively. For each utterance, two speech signals and one noise signal are randomly selected from the 100-hour Librispeech subset \cite{panayotov2015librispeech} and the 100 Non-speech Corpus \cite{web100nonspeech}, respectively. The overlap ratio between the two speakers is uniformly sampled between 0\% and 100\%, and the two speech signals are shifted accordingly and rescaled to a random relative signal-to-noise-ratio(SNR) between 0 and 5 dB. The relative SNR between the sum of the two clean speech power and the noise is randomly sampled between 10 and 20 dB. The transformed signals are then convolved with the room impulse responses simulated by the image method \cite{allen1979image} using the gpuRIR toolbox \cite{diaz2020gpurir}. The length and width of all the rooms are randomly sampled between 3 and 10 meters, and the height is randomly sampled between 2.5 and 4 meters. The reverberation time (T60) is randomly sampled between 0.1 and 0.5 seconds. After convolution, the echoic signals are summed to create the mixture for each microphone.

\subsection{Model configurations}

We follow the standard configuration of DPRNN-TasNet in all models. The total number of DPRNN blocks $M$ in the DPRNN-TasNet is set to 6 in all models. The window size in the waveform encoder and decoder is set to 2~ms (32~samples), and the number of filters in the encoder and decoder is always 128. The input size and hidden size of the LSTM layers in the DPRNN blocks are set to 64 and 128, respectively. 

Note that in the mixed SIMO-SISO design, the SIMO module can contain no DPRNN blocks but simply a single FC layer to generate the $C$ intermediate features. In this case, the SISO module contains all 6 DPRNN blocks similar to the SISO-only design. We exclude this configuration from the SISO-only design as it does not perform iterative separation.

\subsection{Training configurations}

All models are trained for 100 epochs with the Adam optimizer \cite{kingma2014adam} with an initial learning rate of 0.001. Signal-to-noise ratio (SNR) is used as the training objective for all models. The learning rate is decayed by 0.98 for every two epochs. Gradient clipping by a maximum gradient norm of 5 is always applied for proper convergence of DPRNN-based models. Early stopping is applied when no best validation model is found for 10 consecutive epochs. Auxiliary autoencoding training (A2T) is applied to enhance the robustness on this reverberant separation task \cite{luo2020distortion}.

\section{Results and discussions}
\label{sec:result}
\subsection{Performance of SIMO-only and mixed SIMO-SISO designs}

\begin{table}[!ht]
	\scriptsize
	\centering
	\begin{tabular}{c|c|cccc|c}
		\thline
		\multirow{2}{*}{SIMO blocks} & \multirow{2}{*}{SISO blocks} & \multicolumn{4}{c|}{Overlap ratio (\%)} & \multirow{2}{*}{Average} \\
		\cline{3-6}
		& & $<$25 & 25-50 & 50-75 & $>$75 \\
		\thline
		6 & 0 & 13.9 & 10.0 & 7.2 & 4.8 & 9.0 \\
		\hline
		5 & 1 & 14.0 & 10.1 & 7.3 & 4.9 & 9.1 \\
		4 & 2 & 14.2 & 10.4 & 7.6 & \textbf{5.0} & 9.4 \\
		3 & 3 & 14.4 & 10.5 & 7.6 & \textbf{5.0} & 9.4 \\
		2 & 4 & \textbf{14.6} & \textbf{10.6} & \textbf{7.8} & 4.9 & \textbf{9.5} \\
		1 & 5 & 14.3 & 10.3 & 7.5 & 4.8 & 9.2 \\
		0 & 6 & 13.5 & 9.5 & 6.8 & 4.5 & 8.6 \\
		\thline
	\end{tabular}
	\caption{Separation performance of different configurations in the SIMO-only and mixed SIMO-SISO designs across different overlap ratios between the two speakers. SI-SDR is reported in decibel scale.}
	\label{tab:result}
\end{table}

We start with the results on the SIMO-only and mixed SIMO-SISO designs. Table~\ref{tab:result} presents the separation performance of the models in the two designs across different overlap ratios between the speakers. The first row presents the standard SIMO-only design, which is also the design for the original DPRNN-TasNet. All other rows show the performance of mixed SIMO-SISO design with different numbers of blocks in each module. We first notice that despite the configuration of 0 SIMO blocks, all other SIMO-SISO configurations lead to better performance than the standard SIMO-only design. Moreover, best performance is achieved at the 4-block configuration, and the 2- and 3-block configurations also lead to comparable performance. The worst performance is observed at the 6-block configuration. This indicates that a deeper design in the SISO module is able to improve the performance, while the separation layers in the SIMO module also play an important role. A balance can be found on the arragement of the number of layers in the SIMO and SISO modules, and we empirically observe here that assigning 70\% of the total blocks to the SISO module can be a good configuration.

Another finding from the table is that the performance improvement obtained by the mixed SIMO-SISO design mainly comes from the low-overlap utterances. The performance on the utterance with higher than 75\% overlap ratio is consistent across all configurations, however the performance on utterances with lower than 25\% overlap ratio can vary by 1 dB. This implies that the mixed SIMO-SISO design might be more important for the single-speaker regions. One possible explanation comes from the role of the output layer of the SIMO module. In the standard SIMO-only design where the output FC layer estimates the $C$ masks, the values for the masks have to be zero for inactive speakers in the single-speaker regions. Since the $C$ output heads in the FC layer all receive a same feature from the output of the second last layer in the SIMO module, the estimation of the $C$ masks not only requires the feature to be linearly separable in the latent space defined by the parameters of the FC layer, but also forces the same set of parameters to be able to reconstruct salient and silent regions across different regions. This may introduce difficulties on the optimization and put extra requirements on the feature dimension in order to achieve such constraints. Using a deeper SISO module removes the second constraint on the single-speaker regions and does not harm the first contraint on the separability. As more and more recent models consider data distributions with partially-overlap utterances \cite{gu2019end, cosentino2020librimix, chen2020continuous}, such mixed SIMO-SISO design should be more practical and beneficial than the standard designs.

\subsection{Performance of SISO-only design}

\begin{table}[!ht]
	\scriptsize
	\centering
	\begin{tabular}{c|c|cccc|c}
		\thline
		Encoder & Decoder & \multicolumn{4}{c|}{Overlap ratio (\%)} & \multirow{2}{*}{Average} \\
		\cline{3-6}
		blocks & blocks & $<$25 & 25-50 & 50-75 & $>$75 \\
		\thline
		1 & 5 & \textbf{14.3} & \textbf{10.3} & 7.3 & \textbf{4.9} & \textbf{9.3} \\
		2 & 4 & 14.2 & 10.2 & \textbf{7.4} & 4.8 & 9.1 \\
		3 & 3 & 14.0 & 10.0 & 7.1 & 4.4 & 8.9 \\
		4 & 2 & 13.4 & 9.3 & 6.5 & 3.8 & 8.3 \\
		5 & 1 & 13.0 & 8.9 & 6.2 & 3.3 & 7.9 \\
		\thline
	\end{tabular}
	\caption{Separation performance of different configurations in the SISO-only design across different overlap ratios between the two speakers. SI-SDR is reported in decibel scale.}
	\label{tab:result-single}
\end{table}

We then provide the experiment results on different configurations in the SISO-only design. Table~\ref{tab:result-single} shows the separation performance on different numbers of encoding and decoding layers described in Section~\ref{sec:separate}. The performance is getting consistently worse as the number of decoder blocks decreases, implying that the model capacity in the decoder blocks need to be large enough in such iterative separation scheme. The best performance, on the other hand, is still slightly better than the standard SIMO-only design, especially on the low-overlap utterances. This matches our discussions in the previous section about the importance of deeper architectures for the single-speaker regions in the mixture. 

The results provide another perspective on the role of the SIMO separation layers in a BSS network and rise new questions. If the SIMO module is not even necessary for successful separation, then what are the roles of the separation layers in a separation network? How are the speaker-dependent features, including speaker identity and contents of the context, separated by the SISO-only models? If unbiased speech extraction can replace speech separation, can we find a unified SISO framework for both GSS and BSS? Such questions may open new discussions on our understanding of separation networks and motivate new design paradigms for new architectures.

\section{Conclusion}
\label{sec:conclusion}
In this paper, we revisited the roles of \textit{separation layers} in speech separation networks. Any separation network can be split into SIMO and SISO modules, where the SIMO modules generating more outputs than input were defined as the separation layers in the entire network. We explored three model configurations with a same backbone: the SIMO-only design, the mixed SIMO-SISO design, and the iterative SISO-only design. Experiment results on various configurations showed that although almost all existing separation systems are SIMO-only, the mixed SIMO-SISO design can improve the separation performance especially on low-overlap utterances. The SISO-only design also achieved slightly better performance than the standard SIMO-only design, challenging the role of the SIMO separation layers in a speech separation system. The results allowed us to rethink the problem formulation of speech separation and the design paradigm for separation systems. 

\section{Acknowledgments}
This work was funded by a grant from the National Institute of Health, NIDCD, DC014279; a National Science Foundation CAREER Award; and the Pew Charitable Trusts. This work was also carried out during the 2020 Jelinek Memorial Summer Workshop on Speech and Language Technologies at Johns Hopkins University (JSALT 2020), supported with unrestricted gifts from Microsoft (Research and Azure), Amazon (Alexa and AWS), and Google.

\bibliographystyle{IEEEtran}
\bibliography{refs}

\begin{thebibliography}{10}
\providecommand{\url}[1]{#1}
\csname url@samestyle\endcsname
\providecommand{\newblock}{\relax}
\providecommand{\bibinfo}[2]{#2}
\providecommand{\BIBentrySTDinterwordspacing}{\spaceskip=0pt\relax}
\providecommand{\BIBentryALTinterwordstretchfactor}{4}
\providecommand{\BIBentryALTinterwordspacing}{\spaceskip=\fontdimen2\font plus
\BIBentryALTinterwordstretchfactor\fontdimen3\font minus
  \fontdimen4\font\relax}
\providecommand{\BIBforeignlanguage}[2]{{%
\expandafter\ifx\csname l@#1\endcsname\relax
\typeout{** WARNING: IEEEtran.bst: No hyphenation pattern has been}%
\typeout{** loaded for the language `#1'. Using the pattern for}%
\typeout{** the default language instead.}%
\else
\language=\csname l@#1\endcsname
\fi
#2}}
\providecommand{\BIBdecl}{\relax}
\BIBdecl

\bibitem{hershey2016deep}
J.~R. Hershey, Z.~Chen, J.~Le~Roux, and S.~Watanabe, ``Deep clustering:
  Discriminative embeddings for segmentation and separation,'' in
  \emph{Acoustics, Speech and Signal Processing (ICASSP), 2016 IEEE
  International Conference on}.\hskip 1em plus 0.5em minus 0.4em\relax IEEE,
  2016, pp. 31--35.

\bibitem{yu2017permutation}
D.~Yu, M.~Kolb{\ae}k, Z.-H. Tan, and J.~Jensen, ``Permutation invariant
  training of deep models for speaker-independent multi-talker speech
  separation,'' in \emph{Acoustics, Speech and Signal Processing (ICASSP), 2017
  IEEE International Conference on}.\hskip 1em plus 0.5em minus 0.4em\relax
  IEEE, 2017, pp. 241--245.

\bibitem{luo2018tasnet}
Y.~Luo and N.~Mesgarani, ``Tas{N}et: time-domain audio separation network for
  real-time, single-channel speech separation,'' in \emph{Acoustics, Speech and
  Signal Processing (ICASSP), 2018 IEEE International Conference on}.\hskip 1em
  plus 0.5em minus 0.4em\relax IEEE, 2018.

\bibitem{wang2018alternative}
Z.-Q. Wang, J.~Le~Roux, and J.~R. Hershey, ``Alternative objective functions
  for deep clustering,'' in \emph{Acoustics, Speech and Signal Processing
  (ICASSP), 2018 IEEE International Conference on}, 2018.

\bibitem{le2019sdr}
J.~Le~Roux, S.~Wisdom, H.~Erdogan, and J.~R. Hershey, ``{SDR}--half-baked or
  well done?'' in \emph{Acoustics, Speech and Signal Processing (ICASSP), 2019
  IEEE International Conference on}.\hskip 1em plus 0.5em minus 0.4em\relax
  IEEE, 2019, pp. 626--630.

\bibitem{drude2019sms}
L.~Drude, J.~Heitkaemper, C.~Boeddeker, and R.~Haeb-Umbach, ``{SMS}-{WSJ}:
  Database, performance measures, and baseline recipe for multi-channel source
  separation and recognition,'' \emph{arXiv preprint arXiv:1910.13934}, 2019.

\bibitem{wichern2019wham}
G.~Wichern, J.~Antognini, M.~Flynn, L.~R. Zhu, E.~McQuinn, D.~Crow, E.~Manilow,
  and J.~L. Roux, ``Wham!: Extending speech separation to noisy environments,''
  \emph{arXiv preprint arXiv:1907.01160}, 2019.

\bibitem{cosentino2020librimix}
J.~Cosentino, M.~Pariente, S.~Cornell, A.~Deleforge, and E.~Vincent,
  ``Librimix: An open-source dataset for generalizable speech separation,''
  \emph{arXiv preprint arXiv:2005.11262}, 2020.

\bibitem{chen2020continuous}
Z.~Chen, T.~Yoshioka, L.~Lu, T.~Zhou, Z.~Meng, Y.~Luo, J.~Wu, X.~Xiao, and
  J.~Li, ``Continuous speech separation: Dataset and analysis,'' in
  \emph{Acoustics, Speech and Signal Processing (ICASSP), 2020 IEEE
  International Conference on}.\hskip 1em plus 0.5em minus 0.4em\relax IEEE,
  2020, pp. 7284--7288.

\bibitem{vincent2014blind}
E.~Vincent, N.~Bertin, R.~Gribonval, and F.~Bimbot, ``From blind to guided
  audio source separation: How models and side information can improve the
  separation of sound,'' \emph{IEEE Signal Processing Magazine}, vol.~31,
  no.~3, pp. 107--115, 2014.

\bibitem{wang2018voicefilter}
Q.~Wang, H.~Muckenhirn, K.~Wilson, P.~Sridhar, Z.~Wu, J.~Hershey, R.~A.
  Saurous, R.~J. Weiss, Y.~Jia, and I.~L. Moreno, ``Voicefilter: Targeted voice
  separation by speaker-conditioned spectrogram masking,'' \emph{arXiv preprint
  arXiv:1810.04826}, 2018.

\bibitem{chen2018multi}
Z.~Chen, X.~Xiao, T.~Yoshioka, H.~Erdogan, J.~Li, and Y.~Gong, ``Multi-channel
  overlapped speech recognition with location guided speech extraction
  network,'' in \emph{2018 IEEE Spoken Language Technology Workshop
  (SLT)}.\hskip 1em plus 0.5em minus 0.4em\relax IEEE, 2018, pp. 558--565.

\bibitem{kanda2019guided}
N.~Kanda, C.~Boeddeker, J.~Heitkaemper, Y.~Fujita, S.~Horiguchi, K.~Nagamatsu,
  and R.~Haeb-Umbach, ``Guided source separation meets a strong {ASR} backend:
  Hitachi/paderborn university joint investigation for dinner party asr,''
  \emph{arXiv preprint arXiv:1905.12230}, 2019.

\bibitem{vzmolikova2019speakerbeam}
K.~{\v{Z}}mol{\'\i}kov{\'a}, M.~Delcroix, K.~Kinoshita, T.~Ochiai, T.~Nakatani,
  L.~Burget, and J.~{\v{C}}ernock{\`y}, ``Speakerbeam: Speaker aware neural
  network for target speaker extraction in speech mixtures,'' \emph{IEEE
  Journal of Selected Topics in Signal Processing}, vol.~13, no.~4, pp.
  800--814, 2019.

\bibitem{xu2019time}
C.~Xu, W.~Rao, E.~S. Chng, and H.~Li, ``Time-domain speaker extraction
  network,'' in \emph{Automatic Speech Recognition and Understanding (ASRU),
  2019 IEEE Workshop on}.\hskip 1em plus 0.5em minus 0.4em\relax IEEE, 2019,
  pp. 327--334.

\bibitem{kolbaek2017multitalker}
M.~Kolb{\ae}k, D.~Yu, Z.-H. Tan, and J.~Jensen, ``Multitalker speech separation
  with utterance-level permutation invariant training of deep recurrent neural
  networks,'' \emph{IEEE/ACM Transactions on Audio, Speech, and Language
  Processing (TASLP)}, vol.~25, no.~10, pp. 1901--1913, 2017.

\bibitem{luo2019conv}
Y.~Luo and N.~Mesgarani, ``Conv-{T}as{N}et: Surpassing ideal time--frequency
  magnitude masking for speech separation,'' \emph{IEEE/ACM Transactions on
  Audio, Speech, and Language Processing (TASLP)}, vol.~27, no.~8, pp.
  1256--1266, 2019.

\bibitem{stoller2018wave}
D.~Stoller, S.~Ewert, and S.~Dixon, ``Wave-{U}-{N}et: A multi-scale neural
  network for end-to-end audio source separation,'' \emph{arXiv preprint
  arXiv:1806.03185}, 2018.

\bibitem{zeghidour2020wavesplit}
N.~Zeghidour and D.~Grangier, ``Wavesplit: End-to-end speech separation by
  speaker clustering,'' \emph{arXiv preprint arXiv:2002.08933}, 2020.

\bibitem{nachmani2020voice}
E.~Nachmani, Y.~Adi, and L.~Wolf, ``Voice separation with an unknown number of
  multiple speakers,'' \emph{arXiv preprint arXiv:2003.01531}, 2020.

\bibitem{takahashi2019recursive}
N.~Takahashi, S.~Parthasaarathy, N.~Goswami, and Y.~Mitsufuji, ``Recursive
  speech separation for unknown number of speakers,'' \emph{Proc. Interspeech},
  pp. 1348--1352, 2019.

\bibitem{chang2019mimo}
X.~Chang, W.~Zhang, Y.~Qian, J.~Le~Roux, and S.~Watanabe, ``{MIMO}-speech:
  End-to-end multi-channel multi-speaker speech recognition,'' in
  \emph{Automatic Speech Recognition and Understanding (ASRU), 2019 IEEE
  Workshop on}.\hskip 1em plus 0.5em minus 0.4em\relax IEEE, 2019, pp.
  237--244.

\bibitem{delfarah2019deep}
M.~Delfarah and D.~Wang, ``Deep learning for talker-dependent reverberant
  speaker separation: An empirical study,'' \emph{IEEE/ACM Transactions on
  Audio, Speech, and Language Processing}, vol.~27, no.~11, pp. 1839--1848,
  2019.

\bibitem{yoshioka2019low}
T.~Yoshioka, Z.~Chen, C.~Liu, X.~Xiao, H.~Erdogan, and D.~Dimitriadis,
  ``Low-latency speaker-independent continuous speech separation,'' in
  \emph{Acoustics, Speech and Signal Processing (ICASSP), 2019 IEEE
  International Conference on}.\hskip 1em plus 0.5em minus 0.4em\relax IEEE,
  2019, pp. 6980--6984.

\bibitem{luo2019dual}
Y.~Luo, Z.~Chen, and T.~Yoshioka, ``Dual-path {RNN}: efficient long sequence
  modeling for time-domain single-channel speech separation,'' \emph{arXiv
  preprint arXiv:1910.06379}, 2019.

\bibitem{kanda2020serialized}
N.~Kanda, Y.~Gaur, X.~Wang, Z.~Meng, and T.~Yoshioka, ``Serialized output
  training for end-to-end overlapped speech recognition,'' \emph{arXiv preprint
  arXiv:2003.12687}, 2020.

\bibitem{luo2020end}
Y.~Luo, Z.~Chen, N.~Mesgarani, and T.~Yoshioka, ``End-to-end microphone
  permutation and number invariant multi-channel speech separation,'' in
  \emph{Acoustics, Speech and Signal Processing (ICASSP), 2020 IEEE
  International Conference on}.\hskip 1em plus 0.5em minus 0.4em\relax IEEE,
  2020, pp. 6394--6398.

\bibitem{panayotov2015librispeech}
V.~Panayotov, G.~Chen, D.~Povey, and S.~Khudanpur, ``Librispeech: an {ASR}
  corpus based on public domain audio books,'' in \emph{Acoustics, Speech and
  Signal Processing (ICASSP), 2015 IEEE International Conference on}.\hskip 1em
  plus 0.5em minus 0.4em\relax IEEE, 2015, pp. 5206--5210.

\bibitem{web100nonspeech}
G.~Hu, ``100 {N}onspeech {S}ounds,'' {\small
  \url{http://web.cse.ohio-state.edu/pnl/corpus/HuNonspeech/HuCorpus.html}}.

\bibitem{allen1979image}
J.~B. Allen and D.~A. Berkley, ``Image method for efficiently simulating
  small-room acoustics,'' \emph{The Journal of the Acoustical Society of
  America}, vol.~65, no.~4, pp. 943--950, 1979.

\bibitem{diaz2020gpurir}
D.~Diaz-Guerra, A.~Miguel, and J.~R. Beltran, ``gpu{RIR}: A python library for
  room impulse response simulation with gpu acceleration,'' \emph{Multimedia
  Tools and Applications}, pp. 1--19, 2020.

\bibitem{kingma2014adam}
D.~Kingma and J.~Ba, ``Adam: A method for stochastic optimization,''
  \emph{arXiv preprint arXiv:1412.6980}, 2014.

\bibitem{luo2020distortion}
Y.~Luo, C.~Han, and N.~Mesgarani, ``Distortion-controlled training for
  end-to-end reverberant speech separation with auxiliary autoencoding loss,''
  in \emph{2021 IEEE Spoken Language Technology Workshop (SLT)}.\hskip 1em plus
  0.5em minus 0.4em\relax IEEE, 2021.

\bibitem{gu2019end}
R.~Gu, J.~Wu, S.-X. Zhang, L.~Chen, Y.~Xu, M.~Yu, D.~Su, Y.~Zou, and D.~Yu,
  ``End-to-end multi-channel speech separation,'' \emph{arXiv preprint
  arXiv:1905.06286}, 2019.

\end{thebibliography}

\end{document}